\documentclass[useAMS,usenatbib]{mn2e}
\usepackage{color}
\usepackage{graphicx}
\usepackage{mathrsfs}
\usepackage{epsfig}
\usepackage{natbib}
\usepackage{color}
\usepackage{float}
\usepackage{amsmath}
\usepackage{times}
\usepackage{upgreek}
\usepackage[varg]{txfonts}
\usepackage{enumerate}
\usepackage{verbatim}
\usepackage{booktabs}
\usepackage{multirow}
\usepackage{amssymb}
\usepackage{placeins}
\usepackage{array,longtable}
\usepackage{appendix}

\newcommand{\toiii}{[O\,{\sc iii}]$_{\lambda5007}$}

\newcommand{\orat}{[O\,{\sc iii}]/[O\,{\sc ii}]}

\title[Two blueberries]{The Dawn Light of Blueberry Galaxies: Spectroscopic and Photometric Studies of two Starburst Dwarf Galaxies}
\author[Rong et al.]{Yu Rong$^{1,2,3}$\thanks{E-mail: rongyuastrophysics@gmail.com}, Huan Yang$^{4}$, Hong-Xin Zhang$^{4}$, Thomas H. Puzia$^{1}$, Igor V. Chilingarian$^{5,6}$, \and Paul Eigenthaler$^{1,2}$, Sangeeta Malhotra$^{7}$, James E. Rhoads$^{7}$, Junxian Wang$^{4}$, \and Yasna Ordens-Brice\~no$^{1,8}$, Evelyn Johnston$^{1}$\\
$^{1}$Instituto de Astrof\'isica, Pontificia Universidad Cat\'olica de Chile, Av. Vicu\~na Mackenna 4860, Macul, Santiago, Chile\\
$^{2}$Chinese Academy of Sciences South America Center for Astronomy and China-Chile Joint Center for Astronomy, Camino El Observatorio 1515,\\ Las Condes, Santiago, Chile\\
$^{3}$National Astronomical Observatories, Chinese Academy of Sciences, 20A Datun Road, Chaoyang District, Beijing 100012, China\\
$^{4}$CAS Key Laboratory for Research in Galaxies and Cosmology, Department of Astronomy, University of Science and Technology of China, China\\
$^{5}$Smithsonian Astrophysical Observatory, 60 Garden St. MS09, Cambridge, MA, 02138, USA\\
$^{6}$Sternberg Astronomical Institute, M.V. Lomonosov Moscow State University, 13 Universitetsky prospect, Moscow, 119991, Russia\\
$^{7}$Arizona State University, School of Earth and Space Exploration, USA\\
$^{8}$Astronomisches Rechen-Institut, Zentrum f\"ur Astronomie der Universit\"at Heidelberg, M\"onchhofstra$\beta$e 12-14, D-69120 Heidelberg, Germany
}

\begin{document}
\maketitle

\begin{abstract}
	
A population of so-called ``blueberry" starbursting dwarf galaxies with extremely blue colors, low-metallicities, and enormous ionization ratios, has recently been found by \cite{Yang17}. Yet we still do not know their detailed properties, such as morphologies, AGN occupations, massive star contents, infrared emission, dust properties, etc. As a pilot study of the blueberries, we investigate the spectroscopic and photometric properties of two blueberry candidates, RGG\,B and RGG\,5, for which Hubble Space Telescope high-resolution images are available.~We find that RGG\,B and RGG\,5 perhaps are likely to be two merging dwarf galaxy systems.~RGG\,B may have a close merging companion; yet the current evidence still cannot exclude the possibility that RGG~B is just disturbed by in-situ star formation through, e.g., outflows, rather than undergoing a merger. RGG\,5 presents stellar shells in the outskirt which can be the powerful evidence of galaxy merging. We also find that, all of the blueberries, including RGG\,B and RGG\,5, are located close to the theoretical maximum-starburst-line in the BPT diagram, have very high ionization parameters, and relatively low hardness ionizing radiation fields, exhibit nitrogen overabundances, and show extremely red mid-IR colors, and reside in the so-called ``ULIRGs/LINERs/Obscured AGN'' region. The blueberry galaxies may not harbor AGN.

\end{abstract}
\section{Introduction} \label{sec:intro}

According to the different photometry and spectroscopy properties, dwarf galaxies can be classified as many intriguing sub-samples, e.g., the blue compact dwarf galaxies \citep[BCDs; e.g.,][]{Zwicky71,Thuan81,Gil03}, ultra-compact dwarfs \citep[e.g.,][]{Drinkwater00,Corbin06}, H\,{\sc ii} galaxies \citep[e.g.,][]{Terlevich81,Melnick17}, extreme metal-poor galaxies \citep[e.g.,][]{Guseva17,Sanchez16}, ultra-diffuse galaxies \citep[e.g.,][]{vanDokkum15,Rong17a}, dwarfs purportedly lacking dark matter \citep{vanDokkum18}, suggesting an intricate and complex multitude of dwarf galaxy evolution senarios.

In the census of the SDSS low-redshift ($z\!<\!0.05$) galaxies, a population of galaxies with low-masses ($\sim\!10^{6.5\--8.0}\ M_{\odot}$), low-metallicities ($7 < 12 + \log({\rm{O/H}}) < 8$), and extremely blue colors (i.e.~$g-r<-0.5$~mag and $r-i<1.0$~mag), has recently been classified as `blueberry' galaxies \citep[hereafter Y17]{Yang17}. Their extremely blue colors are attributed to the strong \toiii\ emission lines with equivalent widths of EW(\toiii)~$\!\gtrsim\!800$\,\AA.~These blueberries are starburst dwarf galaxies with very small sizes ($\lesssim\!1$\,kpc), high specific star-formation rates (sSFRs), and enormous ionization ratios ([O\,{\sc iii}]/[O\,{\sc ii}]~$\!\approx\!10-60$). The properties of blueberries are similar to those of the typical green peas, Ly$\alpha$ galaxies, or genuine H\,{\sc ii} regions, though their stellar masses are only 1\%\--10\% of green peas and Ly$\alpha$ galaxies \citep[e.g.,][]{Finkelstein09}, and might constitute the best local analogs of the lowest-mass starbursts at high redshifts, soon to be observed with JWST \citep[e.g.,][]{Renzini17}.

In their SDSS images, several Y17 blueberry galaxies are not a precise point-spread-function (PSF), but present somewhat diffuse outskirts (see Fig.~2 of Y17). For instance, the galaxies with the numbers of Y17-07, 27, 31, 39, 54, and 70 in the Y17 catalog reveal tails/streams in their halo regions; Y17-10 shows a spiral-arm feature; Y17-67, 12, and 06 present the diffuse envelopes/halos. These may be the evidence of undergoing galaxy mergers. Note, the unresolved SDSS images and spectra of the blueberries prevent us to study their morphologies, compositions, dynamics, and properties of star-forming regions in detail.

One tempting question is whether blueberry galaxies contain central active galactic nuclei (AGN).~Although the Y17 blueberry galaxies are classified as BPT star-forming galaxies \citep{Baldwin81}, they are located at the upper-left region of the star-forming sample, very close to the seperation between AGN and star-forming galaxies. The BPT selection may be ineffective for low-metallicity galaxies located in this parameter region \citep{Groves06, Stasinska06}.

Studying the mid-infrared (mid-IR) colors of the blueberries with the ALLWISE catalog \citep{Wright10}, we find that all of the blueberries are very red, probably suggesting these starburst galaxies are dust obscured.~Indeed, the ultraviolet (UV) radiation from the O- and B-type stars in very young star-forming regions can heat the surrounding dust and contribute significantly to the spectral energy distribution (SED) from the mid-IR to far-infrared (FIR) wavelengths.~\cite{Cormier15} investigated the Herschel PACS spectroscopy of low-metallicity dwarf galaxies, and found that the interstellar medium in these galaxies is more porous than in metal-rich galaxies, leading to a larger fraction of the stellar UV radiation heating the dust.~These results are supported by modeling of \cite{Hirashita04}, who explored that both the dust temperature and luminosity are higher in dense, compact, low-metallicity star-forming regions.

Another senario is that the blueberry objects may be newly-born star clusters, just like the ``open clusters'' formed in H\,{\sc ii} regions \citep[e.g.,][]{Dias02, Balser11, Longmore14}. For instance, the BCD, Mrk~930, reveals a very young cluster population with 70\% of the systems formed less than 10~Myr ago \citep{Adamo11b}. The edges of the young star clusters are places for triggered \citep{Elmegreen98} and progressive star formation \citep[e.g.,][]{Carlson07,Walborn02}.~Delayed or triggered star formation processes in dense and dusty regions surrounding the blue clusters could explain a large fraction of massive young stars contributing to the mid-IR spectrum of a cluster a few Myr old.~Moreover, the [O\,{\sc iii}]/H$\beta$ ratio is known as a tracer of the mean ionization level and temperature of the photoionized gas (Baldwin et al. 1981); the high [O\,{\sc iii}]/H$\beta$ ratios of the blueberries ($\sim\!6\!-\!10$) may be an indication of the young star clusters \citep{Adamo11b}. 

So far, we still do not know the possible origins and detailed morphologies of the blueberries, since for the Y17 blueberries, there are no high-resolution images. This work is a pilot study of blueberry galaxies, and we will investigate the properties of two new blueberry candidates which have high spatial-resolution images, RGG\,B and RGG\,5, identified in the dwarf catalog of \cite{Reines13}.~The former shows broad H$\alpha$ emission but narrow-line ratios consistent with the photoionization from H\,{\sc ii} regions \citep{Pustilnik04,Reines13}, and presents a weak Wolf-Rayet (WR) blue bump. The latter does not show any broad-line emission or WR bump. Therefore, the two dwarfs may stand for two different cases of blueberry galaxies. The Hubble Space Telescope (HST) Wide-Field Camera~3 (WFC3) UVIS/IR observations provide high spatial-resolution images to study the compositions and local environments of the two galaxies.

In Section~\ref{sec:2}, we will study the spectroscopic and photometric properties of the two galaxies, and clarify that the two dwarf galaxies are blueberry candidates and have the similar properties to the Y17 blueberry sample galaxies.~In Section~\ref{sec:3}, we compare the two galaxies and discuss some key properties of the two blueberry candidates, in order to unveil some general properties of blueberries.~We summarize our work in Section~\ref{sec:4}.~Throughout this letter, we use ``dex'' to mean the anti-logarithm, i.e.~0.1\,dex=$10^{0.1}$=1.258, and use $\log$ to mean the decadic logarithm, i.e.~$\log_{10}$.

\section{The blueberry galaxies: RGG\,B and RGG\,5} \label{sec:2}

\begin{figure}
\centering
\includegraphics[width=\columnwidth]{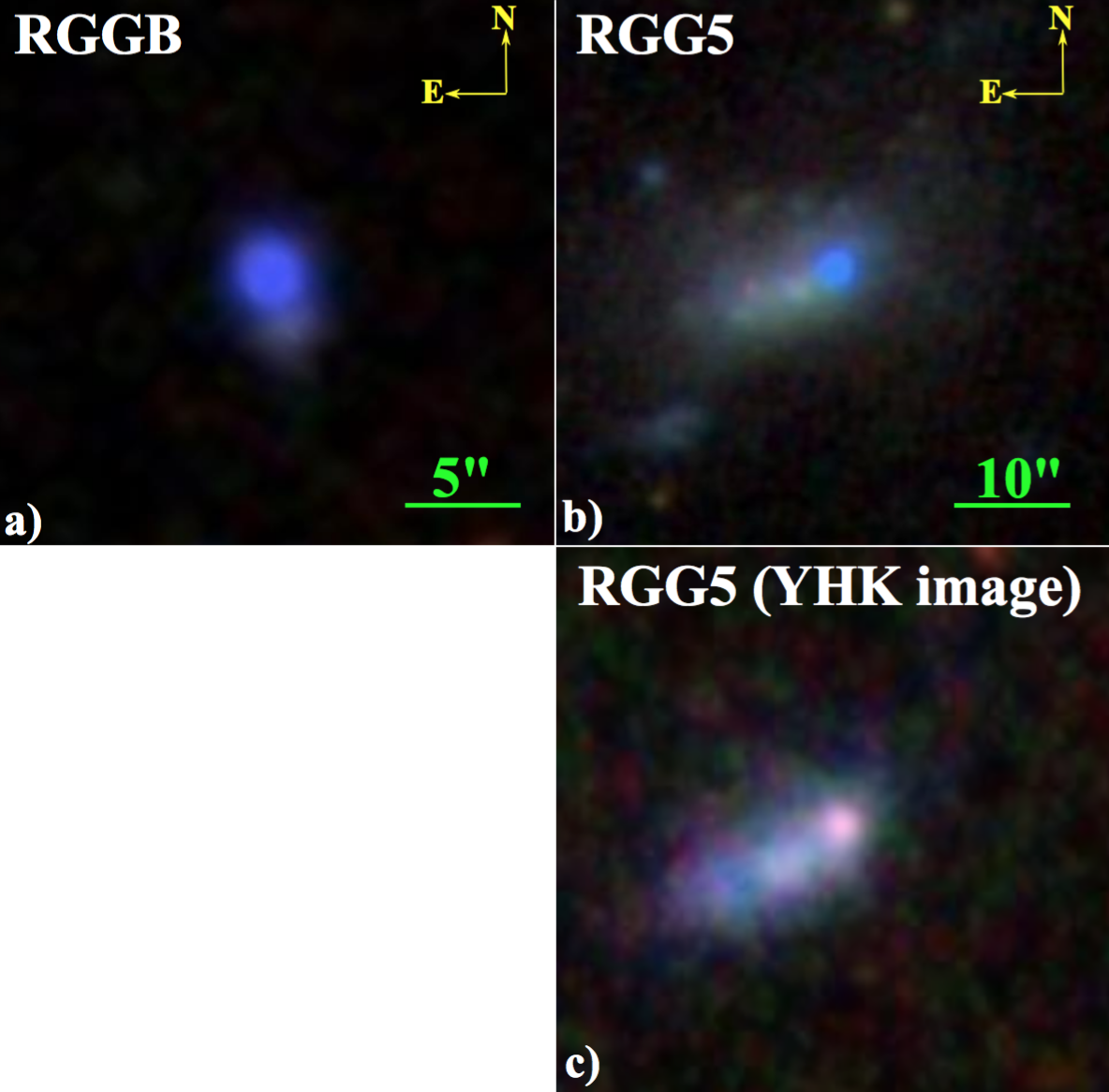}
\caption{The SDSS RGB images for RGG\,B (a) and RGG\,5 (b). Panel~c) shows the NIR YHK image of RGG~5 from UKIDSS.} 
\label{SDSS}
\end{figure}

In this section, we study two dwarf galaxies identified from the catalog of \cite{Reines13}, RGG\,B (SDSS\,J0840+4707) and RGG\,5 (SDSS\,J0823+0313). The two targets are the two bluest galaxies in the dwarf catalog of \cite{Reines13}, which have Hubble Space Telescope high-resolution images; we serendipitously find the two extremely blue dwarfs when we try to study the AGN existence in the local dwarf galaxies with the catalog of \cite{Reines13}. In this section, we will clarify that the two selected galaxies indeed are the blueberry candidates, with similar properties to the Y17 blueberries.

RGG\,B, as shown in panel~a) of Fig.~\ref{SDSS}, is a BCD at the redshift $z\!\approx\!0.0421$ \citep[$D_{\rm{L}}\!\simeq\!187$\,Mpc;][]{Izotov07}, located at RA(J2000)\,=\,08h40m29.9s, DEC(J2000)\,=\,$+47^{\circ}07'10''$, with a low metallicity of $12+\log{\rm (O/H)}\!\simeq\!7.73$ \citep{Perez-Montero11}.~The colors of RGG\,B from the SDSS photometry cModel magnitudes (before foreground extinction correction) are $(g\!-\!r)\!=\!-0.90$~mag, $(g\!-\!i)\!=\!-0.70$~mag, $(g\!-\!u)\!=\!-0.89$~mag, and $(r\!-\!i)\!=\!0.20$~mag, respectively. RGG\,B satisfies the color selection criteria of blueberry galaxies in Y17. It was not selected in the conservative sample of Y17, because of the presence of a tail on the south-west in the SDSS image (as shown in Fig.~\ref{SDSS}), which is slightly beyond the photometric flags (artificial flags; see also the Appendix selection criteria in Y17 for details) of selecting blueberries in Y17. 
However, it is worth noting that, several Y17 blueberry galaxies, e.g., Y17-27, 31, 39, 54, and 70, also present a tail/stream on the edges of galaxies (see Fig.~2 of Y17 for details).

Another blue galaxy, RGG\,5, located at coordinates RA(J2000)\,=\,08h23m34.8s and DEC(J2000)\,=\,$+03^{\circ}13'15.7''$, at a redshift $z\!\approx\!0.0098$ ($D_{\rm{L}}\!\simeq\!42.5$\,Mpc) and is much brighter ($r\!\approx\!16.64$~mag) than the average brightness of the Y17 blueberries. RGG\,5 was not selected in the Y17 sample, because of both the existence of an extended halo region as shown in panel~b) of Fig.~\ref{SDSS}, and because the color (calculated from the SDSS PSF magnitudes) $(g\!-\!r)\!\simeq\!-0.47$~mag is slightly beyond the selection criterion of blueberries, i.e., $(g\!-\!r)\!<\!-0.5$~mag. However, some Y17 blueberries, e.g., Y17-67 and 6, also show a faint halo component.~Note that RGG\,5 has much lower redshift compared with the Y17 blueberries, and thus its faint outskirts may not be detectable if we shift its redshift to $z\sim 0.03\--0.05$.

Their SDSS spectra are shown in Fig.~\ref{spectra}. Here we will compare the spectroscopic properties of the two blueberry candidates with those of the Y17 blueberries.

\subsection{Star Formation Properties}
First, we compare the spectroscopic properties of RGG\,B and RGG\,5 with those of the Y17 blueberries.~We calculate the star formation rates (SFRs) of RGG\,B and RGG\,5 from the H$\alpha$ luminosities \citep{Kennicutt98}, SFR$_{\rm RGG\,B}\!=\!3.11\ M_\odot$/yr and SFR$_{\rm RGG\,5}\!=\!0.35\ M_\odot$/yr.~Note that the estimated SFR of RGG~5 is a lower-limit since the SDSS fiber only covered its blueberry region (H\,{\sc ii} region).~Their stellar masses (${\cal M}_\star$) are obtained from the NASA-Sloan Atlas catalog{\footnote{http://nsatlas.org/data}}), ${\cal M}_{\star,\rm RGG\,B}\!=\!1.4\times 10^8\ M_{\odot}$ and ${\cal M}_{\star,\rm RGG\,5}\!=\!3.73\times 10^8\ M_{\odot}$.~Therefore, analogous to the properties of Y17 blueberry sample, RGG\,B and RGG\,5 show strong star formation activities and are located beyond the ``main-sequence'' in the SFR--${\cal M}_\star$ diagram, compared with most of the SDSS galaxies.~The specific SFRs (sSFRs) of the blueberries are broadly similar to those of the higher-redshift starbursts \citep[e.g.,][]{Atek14,Rodriguez15,Elmegreen17} as shown by the magenta symbols in panel~A of Fig.~\ref{property}, implying that the blueberries may be good local analogs of lowest-mass (${\cal M}_\star\!\lesssim\!10^9\ M_{\odot}$) starbursts at high redshifts.

\subsection{Emission Line Properties}
\begin{figure}
\centering
\includegraphics[width=\columnwidth]{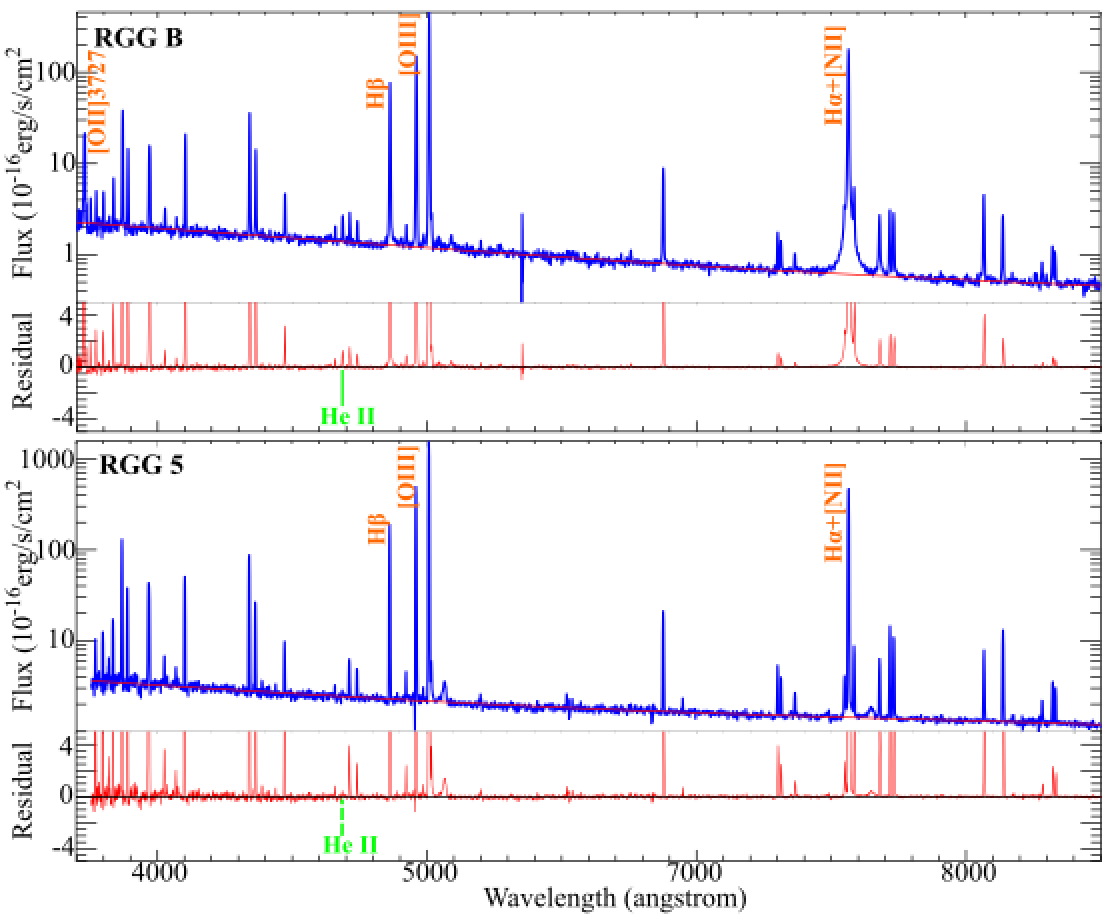}
\caption{The SDSS spectra of RGG\,B (upper) and RGG\,5 (lower), with several emission lines marked (in orange). The red lines are the fitting results to the continuum spectra.~The residuals are shown in the lower sub-panels, where the He\,{\sc ii}$_{\lambda4686}$ line is highlighted in green.~RGG\,B shows strong He\,{\sc ii} emission line, while RGG\,5 exhibits weak He\,{\sc ii} emission.}
\label{spectra}
\end{figure}

The SDSS spectrum of RGG\,B (upper panel of Fig.~\ref{spectra}) shows a very strong \toiii\ emission line with an equivalent width of EW(\toiii)~$\!\approx\!1188$\,\AA, and large \toiii/H$\beta$ ($\log ($\toiii/H$\beta)\!\simeq\!0.80$)
and an oxygen ionization ratio\footnote{In the following, we will refer with the oxygen ionization ratio [O\,{\sc iii}]/[O\,{\sc ii}] to the line ratio ([O\,{\sc iii}]$_{\lambda4959}$ + [O\,{\sc iii}]$_{\lambda5007}$)/[O\,{\sc ii}]$_{\lambda3727}$.} of [O\,{\sc iii}]/[O\,{\sc ii}]~$\!\simeq\!21.4$, 
and reveals broad components in several emission lines including [O\,{\sc iii}]$_{\lambda 4363}$ \citep{Izotov07,Reines13}. 

The EW(\toiii)~$\!\simeq\!2100$\,\AA\ of RGG\,5 is also higher than the selection criterion ($\gtrsim\!800$\,\AA) of the blueberries in Y17.~Here we calculate its oxygen metallicity and \orat\ ratio.~Using the [O\,{\sc iii}]$_{\lambda4363}$, [O\,{\sc iii}]$_{\lambda\lambda4959,5007}$, and [O\,{\sc ii}]$_{\lambda\lambda7320,7330}$ fluxes, we can estimate the metallicity of RGG\,5 by following the $T_{e}$ methodology described in \cite{Izotov06}, deriving $12+\log ({\rm{O/H}})\!\simeq\!8.09$.~RGG\,5 has a relatively higher oxygen abundance, compared with the Y17 blueberries ($12+\log({\rm{O/H}})\!\simeq\!7.1-7.8$).~Since the limited SDSS spectral wavelength-range does not cover the [O\,{\sc ii}]$_{\lambda3727}$ line (as shown in the lower panel of Fig.~\ref{spectra}), its \orat\ is unknown.~However, using the metallicity and equations~(3) and (4) in the study of \cite{Izotov06}, we can predict the flux of [O\,{\sc ii}]$_{\lambda3727}\!\approx\!6.07\times 10^{-14}\ \rm erg/s/cm^2$.~In order to test whether the predicted [O\,{\sc ii}]$_{\lambda3727}$ flux is reasonable, we use another diagnostic for the ionization parameter which has proved effective for BCDs \citep{Stasinska15}, the [Ar\,{\sc iv}]$_{\lambda4740}$/[Ar\,{\sc iii}]$_{\lambda7135}$, and utilize the linear correlation between [O\,{\sc iii}]$_{\lambda 5007}$/[O\,{\sc ii}]$_{\lambda3727}$ and [Ar\,{\sc iv}]$_{\lambda4740}$/[Ar\,{\sc iii}]$_{\lambda7135}$ shown in \cite{Stasinska15} to obtain the flux of [O\,{\sc ii}]$_{\lambda3727}$.~The, thus, calculated flux is $\sim\!6\!\times\!10^{-14}\ \rm erg/s/cm^2$, in excellent agreement with the predicted flux from metallicity.~With these numbers, we derive the oxygen ionization ratio \orat~$\!\simeq\!10.5$.

Therefore, both of the RGG\,B and RGG\,5 have low-metallicities, high-ionization ratios, and strong \toiii emission, which are similar to the properties of the Y17 blueberries.

\begin{figure*}
\centering
\includegraphics[width=\textwidth]{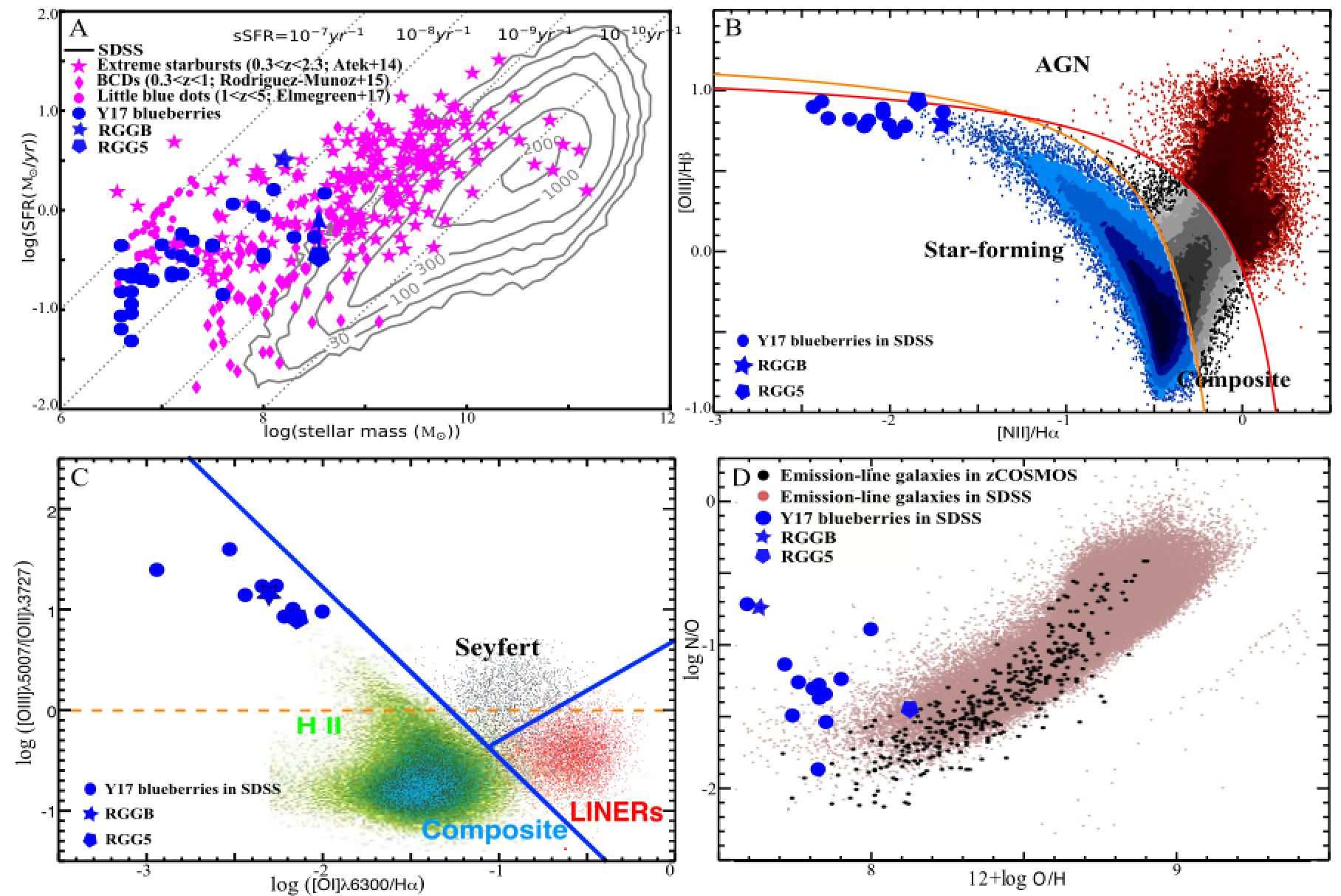}
\caption{In the four panels, the blue points, pentagram, and pentagon mark the Y17 blueberries, RGG\,B, and RGG\,5, respectively. For the four panels, we use the figures from previous literatures and add the points denoting the blueberries in these figures, and compare the properties of the typical star-forming/quescient galaxies with those of the blueberries Panel A, from Y17, shows the SFR versus stellar mass diagram for the blueberry galaxies, extreme starbursts in $0.3<z<2.3$ (magenta stars; Atek et al. 2014), BCDs in $0.3<z<1$ (magenta diamonds; Rodr\'iguez-Mu\~noz et al. 2015), little blue dots in $1<z<5$ (magenta points; Elmegreen et al. 2017), and typical SDSS emission-line galaxies. The arrow indicates that this is the lower-limit of SFR of RGG~5. Panel B is obtained from Trouille et al. (2011); the blue, gray, and red regions in the BPT diagram show the distributions of the galaxies, composites, and AGN/LINERs in SDSS, respectively (Trouille et al. 2011); a darker color explores a higher density. The red and orange curves highlight the extreme starburst line in Kewley et al. (2001) and empirical division between star-forming galaxies and AGN in Kauffmann et al. (2003), respectively. Panel~C, adapted from Kewley et al. (2006), shows the [O\,{\sc iii}]$\lambda5007$/[O\,{\sc ii}]$\lambda3727$ versus [O\,{\sc i}]$\lambda6300$/H$\alpha$ diagnostic diagram for SDSS (green, cyan, red, and black points), Y17 blueberries, and RGG\,B and RGG\,5. The classification scheme shown as blue lines is described in Kewley et al. (2006). The orange dashed line shows the Heckman (1980) LINER line. Panel D, from P\'erez-Montero et al. (2013), shows the N/O versus O/H abundance diagram. The black and brown points denote the emission-line galaxies $z<0.2$ in the SDSS and 20k zCOSMOS surveys, respectively (P\'erez-Montero et al. 2013).
}
\label{property}
\end{figure*}

In the BPT diagram \citep{Baldwin81} as shown in panel~B of Fig.~\ref{property}, RGG\,B and RGG\,5 are located at the similar positions to the Y17 blueberry sample, and very close to the theoretical maximum-starburst-line described in \cite{Kewley01}.~RGG\,5 is slightly beyond the maximum-starburst-line (excess of 0.01~dex); yet considering the 0.1~dex error of the theoretical maximum-starburst-line \citep{Kewley01}, we argue that RGG\,5 may also not contain AGN.

We also plot the diagram of the ionization parameter versus hardness of the ionizing radiation field, i.e., [O\,{\sc iii}]$_{\lambda5007}$/[O\,{\sc ii}]$_{\lambda3727}$ versus [O\,{\sc i}]$_{\lambda6300}$/H$\alpha$, as illustrated in panel~C of Fig.~\ref{property}.~Here we only show the blueberries with the relatively high signal-to-noise ratios (S/N$>5$) of the [O\,{\sc i}]$_{\lambda6300}$ emission lines, to obtain accurate [O\,{\sc i}]$_{\lambda6300}$/H$\alpha$ flux ratio values.~We find that, this particular blueberry population, including RGG\,5 and RGG\,B, are H\,{\sc ii} galaxies with extremely high ionization parameters and relatively low hardness compared with the general population of SDSS galaxies, and cannot be classified as Seyfert galaxies in the diagram \citep{Kewley06}.


The oxygen and nitrogen abundances can be evaluated using the $T_{e}$ method described in \cite{Izotov06}.~The nitrogen-to-oxygen ratio (N/O) versus $12+\log ({\rm{O/H}})$ diagram for the blueberries is then displayed in panel~D of Fig.~\ref{property}.~We find that, compared with the typical emission-line galaxies \citep{Perez-Montero13}, almost all of these low-metallicity starburst blueberries exhibit significant N overabundances which might be produced by nitrogen-enhanced WR star winds \citep[e.g.,][]{Kunth85,Thuan96,Guseva00}, metal-rich outflows, or infall of metal-poor gas \citep{Amorin10,Perez-Montero13}.

The spectroscopic properties of the two dwarf galaxies, RGG\,B and RGG\,5, are broadly similar to those of the Y17 blueberries, and thus can be considered as two blueberry candidates. Therefore, according to the photometric studies of the two blueberry candidates, we can infer, e.g., the morphologies, IR-properties, etc., of the Y17 blueberry population.

\subsection{Photometric Properties}
\subsubsection{Optical and NIR Imaging}

\begin{figure}
\centering
\includegraphics[width=\columnwidth]{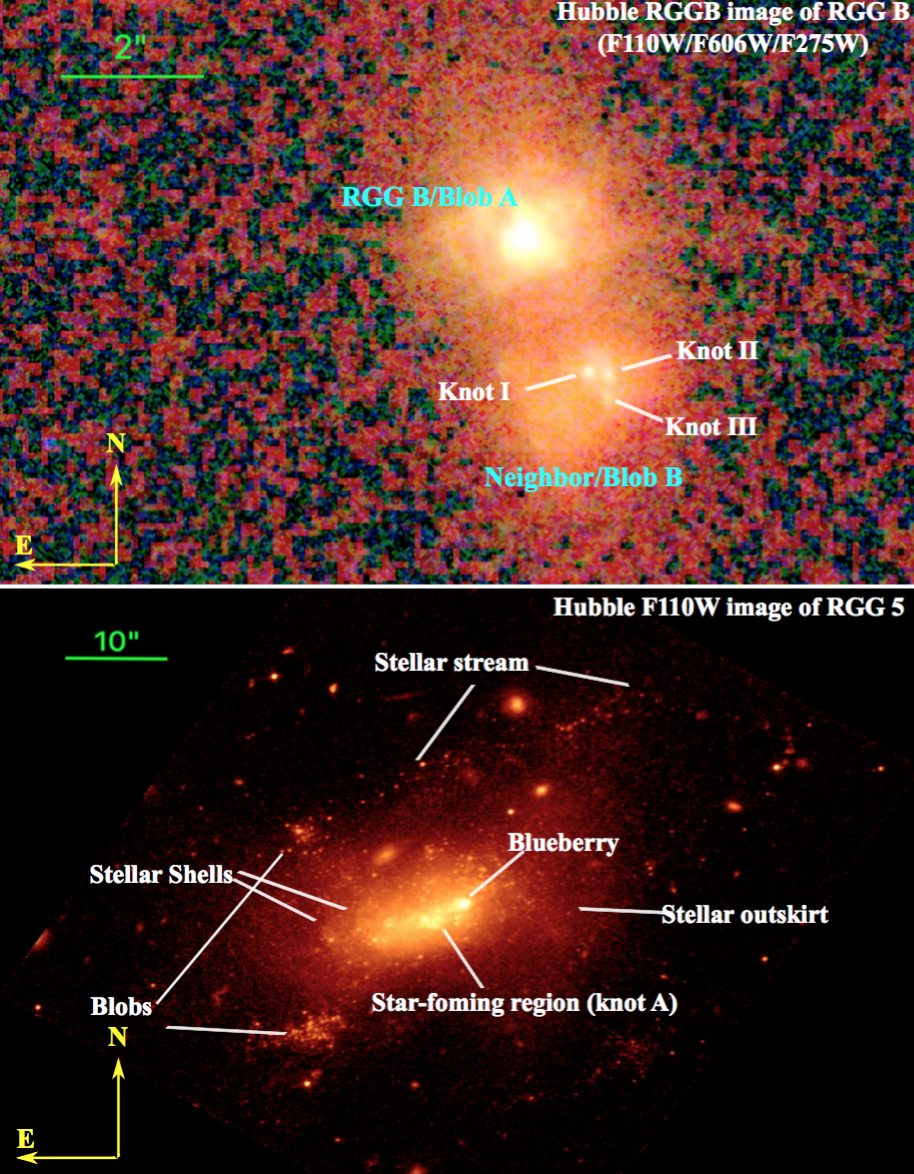}
\caption{High spatial-resolution HST/WFC3 imaging of RGG\,B (upper panel) and RGG\,5 (lower panel).~The image of RGG\,B is a combination of WFC3/UVIS F275W (NUV band), F606W ($V$-band), and WFC3/IR F110W (NIR band) images, while the image of RGG\,5 is a single-passband image taken with the F110W filter.}
\label{HST}
\end{figure}

RGG\,B was observed with the Wide-Field Camera 3 (WFC3) onboard the Hubble Space Telescope (HST) on May 22nd, 2015 (GO-13943, PI: Reines).~In the high-resolution RGB image of RGG\,B, as shown in the upper panel of Fig.~\ref{HST}, it is clear that RGG\,B and its tail in the SDSS image are resolved in two blobs; one possibility is that, the brighter blob~A with the azimuthally more symmetric, radial light distribution might be the galaxy RGG\,B, and the blob~B with several bright knots (knots {\sc i}, {\sc ii}, {\sc iii}) towards the sourth-west of RGG\,B might be its neighbor galaxy.~The irregular galaxy RGG\,B/blob~A has many flocculent/filamentary substrucures, typical of H\,{\sc ii} regions.~The two blobs are about $\sim\!2\arcsec$ apart, which is corresponding to $\lesssim\!2$~kpc at $z\!=\!0.042$.~\cite{Pustilnik04} studied the spectroscopy of the two blobs, with high-S/N spectra from MMT observation, taken on May 1st, 1997, and from SAO RAS 6-m telescope (BTA) observations, taken on February 19th, 2002, whose longslits covered both of the two blobs \citep{Pustilnik04}.~These spectra revealed that the redshift difference between the two blobs is very small with a radial velocity difference of $\lesssim\!60$~km/s.~Therefore, the neighbor/blob~B is very likely to be the merging companion of RGG\,B/blob~A.

The absolute magnitudes (corrected for the foreground extinction) of the two blobs in the three filter bands are listed in Table~\ref{info}; blob~A is much brighter than blob~B.~We also calculate the full-width-at-half-maximum (FWHM) of blob~A in each band, and find that its NUV light, which is produced by the recent star formation and/or AGN, is more concentrated in the central luminous ``nucleus'', while the optical and NIR photons from the relatively-old stellar populations are distributed more diffusely.~Therefore, there is a very dense star-forming region or AGN in the nucleus of blob~A with a color of $({\rm NUV}\!-\!V)\!\simeq \!0.9$~mag.~Blob~B reveals a stellar disk with the S\'ersic index at $\sim\!1$ and several star-forming regions (shown as the three bright knots {\sc i}, {\sc ii}, {\sc iii}) in the RGB image.~The colors of the three knots are $({\rm NUV}\!-\!V)\!\simeq\!0.3$ ({\sc i}), 1.1 ({\sc ii}), and 1.5 ({\sc iii}) mag, respectively, implying age differences between the three knots.~The knot closest to blob~A is youngest, while the farthest one is oldest, in accordance with a merger between RGG\,B/blob~A and its companion/blob~B.

\begin{table} \small
\begin{tabular}{@{}lccc@{}}
\hline
\hline
\multicolumn{4}{c}{RGG\,B} \vspace{1mm}\\
Blob~A & F275W & F606W & F110W \\
Absolute magnitude (mag) & $-17.6$ & $-18.5$ & $-17.6$ \\
FWHM (arcsec) & 0.15 & 0.19 & 0.26 \\
\hline
Blob~B & F275W & F606W & F110W \\
Absolute magnitude (mag) & $-14.7$ & $-16.5$ & $-16.4$ \\
FWHM (arcsec) & 0.85 & 0.95 & 1.13 \\
\hline
Knot~{\sc i}/{\sc ii}/{\sc iii} & F275W & F606W & F110W \\
Absolute magnitude (mag) & \scriptsize{$-12.4$/$-12.3$/$-10.6$} & \scriptsize{$-12.7$/$-13.4$/$-12.1$} & --/--/-- \\
FWHM (arcsec) & 0.10/0.13/0.10 & 0.10/0.25/0.16 & --/--/-- \\
\hline
\hline
\multicolumn{4}{c}{RGG\,5} \vspace{1mm} \\
F110W      & S\'ersic disk & Blueberry & Knot~A \\
Absolute magnitude (mag) & $-17.3$ & $-14.5$ & $-14.4$ \\
S\'ersic index & 0.99 & -- & -- \\
\hline
\hline
\end{tabular}
\caption{Photometric and morphological parameters of RGG\,B and RGG\,5.~For RGG~B, we use a Gaussian profile to fit blob~A, and use a S\'ersic profile plus multi-Gaussian profiles to fit blob~B.~For RGG\,5 we also use a S\'ersic profile to fit the galaxy. ``--'' indicates no measurement.}
\label{info}
\end{table}

However, it is also worth to note that the radial velocity difference between the blob~A and B is only $\lesssim\!60$~km/s according to the long-slit spectroscopic observations \citep[the seeing of the observations is $\sim 2$~arcsec]{Pustilnik04}; this small velocity difference may indicate the, e.g., asymmetric outflows in a galaxy, and thus the two blobs may actually be two star-forming regions in the outflows of one galaxy without merging \citep{Pustilnik04}. Indeed, star formation can significantly disturb the galaxy morphology and reshape the entire galaxy without changing much its internal kinematics, e.g., outflows or galaxy rotation \citep[e.g.,][]{Thuan04,vanZee98}. 
Therefore, although the disturbed overall morphology with the ``double-blob'' structure, together with the very high power of the starburst in RGG~B, suggests a possible recent dwarf merger, yet we still cannot exclude the possibility that RGG~B is just disturbed by the in-situ star formation rather than undergoing a merger.~Detailed future IFU observations will help to resolve this issue.


RGG\,5 was observed with HST WFC3/IR on Jan 27th, 2016 (SNAP-14251, PI: Reines).~In the WFC3/IR F110W image (shown in the lower panel of Fig.~\ref{HST}), RGG\,5 presents a bright knot (corresponding to the blueberry core in the SDSS image, and named as `blueberry' in the bottom panel of Fig.~\ref{HST}), a remarkable star-forming region (knot~A), stellar outskirts, arc-like stellar shells, some blobs which may contain the populations of new-born star clusters, and a stellar-stream/spiral-arm feature straddling across the North.~Particularly, the stellar shells appear aligned along the major axis of RGG\,5, and indeed are the powerful evidence of a merger between two near-equal-mass dwarf galaxies \citep{Paudel17, Pop17}. 

We find that the smooth surface brightness profile of RGG\,5 can be well approximated by a S\'ersic component, by using the iterative fitting method described in \cite{Eigenthaler18}; the normalized chi-square of fitting is $\sim 10^{-8}$, and the residual image only includes the blueberry core, star-forming hotspots (e.g., knot~A, blobs), and some background galaxy interlopers\footnote{We have also tried double-S\'ersic fitting to compare with the one-S\'ersic fitting result, and find that the residual image and normalized chi-square are very similar to those of the one-S\'ersic fitting results. We, therefore, use the one-S\'ersic fitting for simplification.}. The NIR magnitudes of the blueberry core and star-forming knot~A can thus be obtained from the residual image, as listed in Table~\ref{info}; we find that they have similar absolute F110W magnitudes $\sim -14.5$~mag.

Compared with RGG\,B, RGG\,5 does not reveal many filamentary substructures, and is relatively redder, suggesting that RGG\,5 may be in a slightly older merging stage than RGG\,B. A significant fraction of young star clusters (YSCs) shows a clear signature of a flux excess in the NIR band \citep{Adamo10,Adamo11a}, primarily shown as the luminous point-like sources in the blueberry core, star-forming region knot~A, hotspots around the blueberry and knot~A, blobs, and stellar stream. Particularly for the blueberry core, it exhibits higher flux densities in the NIR passband compared with those in the optical passband regardless of the emission-line fluxes{\footnote{http://rcsed.sai.msu.ru/catalog/587732577221083492}} (e.g., strong [O\,{\sc iii}], H$\alpha$, etc.), and its YHK-combined image from the UKIRT Infrared Deep Sky Survey \citep[UKIDSS;][]{Lawrence07} reveals a ``pink'' color \citep[as shown in panel c of Fig.~\ref{SDSS};][]{Chilingarian12,Chilingarian17}, compared with the blue color of the outskirt, suggesting that its $K$-band flux is relatively brighter than the $Y$-band flux. 
It may indicate a large fraction of red supergiants stars in the stellar population in the blueberry core, as the stellar atmospheres of young red supergiants are relatively cool \citep[e.g.][]{Gustafsson08}, and hence have a dominant light output in NIR.~This would imply very young stellar population ages of a few Myr.~Alternatively, thermally pulsing AGB phase transition stars provide also significant NIR flux output \citep[e.g.][]{Capozzi16}, which would indicate stellar population ages within $\sim\!0.2\!-\!2$\, Gyr \citep{Maraston05}.~However, the high-ionization features in the blueberry spectra favor the red supergiant, and thus, younger stellar population picture.

\subsubsection{Photometric mid-IR Properties}

\begin{figure}
\centering
\includegraphics[width=\columnwidth]{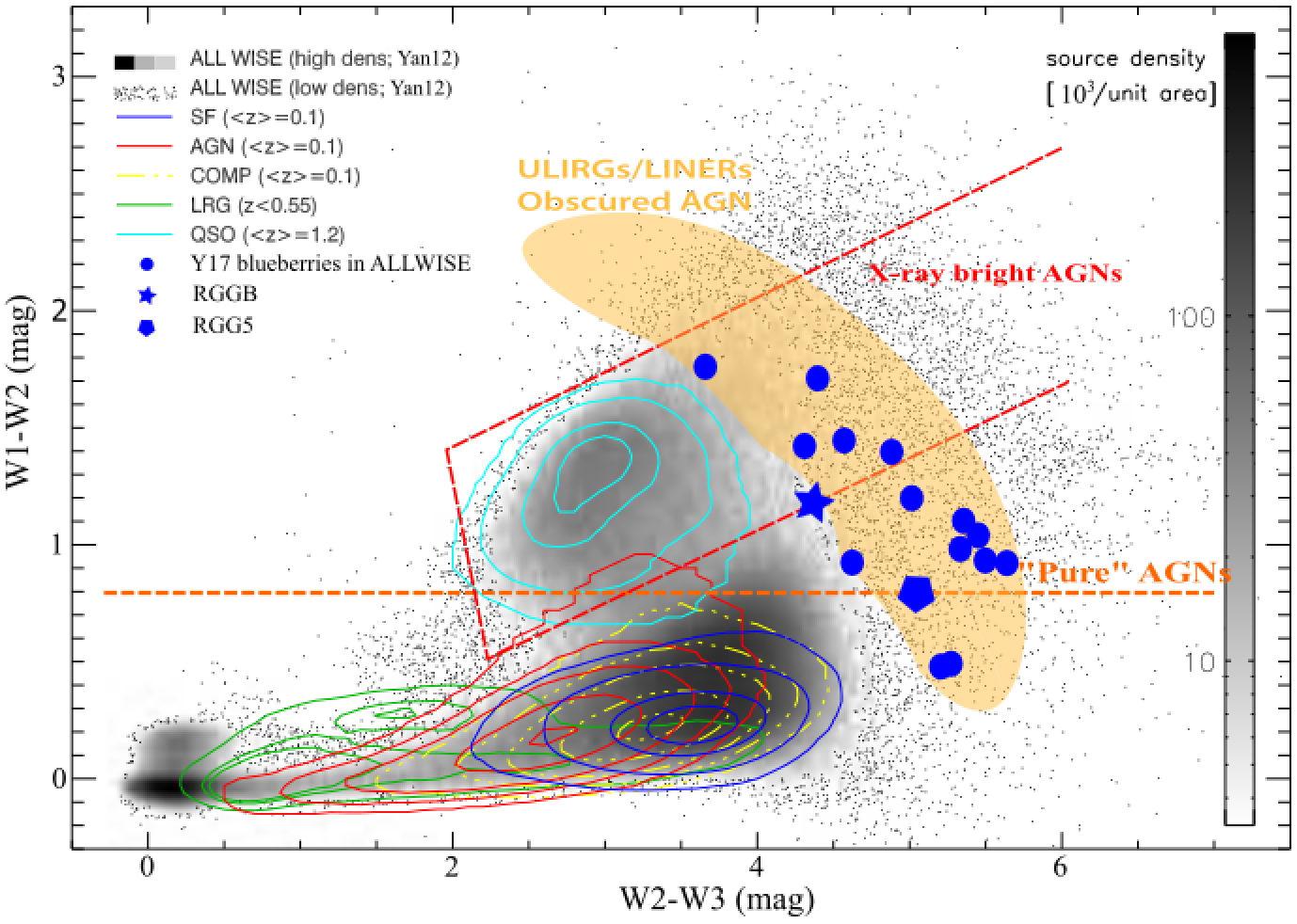}
\caption{Color-color distribution of WISE sources from the 12~$\mu$m-selected W1/2/3 sample; this figure is obtained from Yan et al. (2013), and we add the blueberries (blue points) including RGG\,B (blue pentagram) and RGG\,5 (blue pentagon) in the figure. In order to illustrate the large range in source density in color-color space, the grey-scale and individual points are used to show the high and low source density region, respectively (see Yan et al. (2013) for details). The color contours indicate the concentrations of distributions for the SDSS DR7 sources classified as star forming galaxies (SF, blue), seyfert AGN (red), BPT composite systems (yellow), luminous red galaxies (LRGs; green), and bright quasars (QSOs; cyan).The orange dashed line splits the so-called ``pure'' AGN proposed in Stern et al. (2012). The light orange component coarsely highlights the ultra-luminous infrared galaxy (ULIRGs)/LINERs/Obscured AGN region sketched in Lake et al. (2012). The red dashed wedge brackets the X-ray-bright AGN as clarified in Banfield et al. 2015. The blue points, pentagram, and pentagon mark the Y17 blueberries, RGG\,B, and RGG\,5, respectively.
}
\label{IR}
\end{figure}

In Fig.~\ref{IR} (adapted from Yan et al. (2013)), we show the mid-IR colors, W1-W2 and W2-W3, of RGG\,B and RGG\,5 retrieved from the {\sc AllWISE} Source Catalog \citep{Wright10}.~Both galaxies are extremely red; most of the blueberries are even redder than the so-called ``pure AGN'' line \citep{Stern12}. 
However, note that dwarf galaxies with extreme star formation are capable of heating dust to temperatures producing mid-IR colors W1-W2~$\!>\!0.8$\,mag \citep[e.g.,][]{Stern12}, and this single color cut (W1-W2=0.8) alone should not be used to select AGN in dwarf galaxies \citep{Hainline16}.~We find that the Y17 blueberries and RGG\,B and RGG\,5 coarsely reside in the ``ultra-luminous-infrared galaxy (ULIRGs)/LINERs/Obscured AGN'' region \citep{Lake12} in the color-color diagram, different from the region for the typical starburst galaxies.

\vspace{2mm}
In summary, RGG\,B and RGG\,5 are two blueberry candidates with similar properties as the Y17 blueberries. As a pilot study we analyze the photometry and spectroscopy of these two galaxies as a pathway to the general properties of the blueberry galaxy population as a whole.

\section{Discussion of other properties} \label{sec:3}

\subsection{The possible presence of an AGN?}
Do RGG\,B and RGG\,5 contain AGN? Although Chandra observation of RGG\,B reveals a low X-ray luminosity \citep[$L_{2-10{\rm{keV}}}<10^{40.2}\ \rm{erg/s}$;][]{Baldassare17}, some evidence (e.g., a luminous core in the galaxy) may also agree with the existence of an obscured low-metallicity AGN; particularly, RGG\,B exhibits broad-line emission, e.g.,~broad H$\alpha$, which may originate from the broad-line region of an AGN.~However, here we argue that the AGN may not exist in RGG\,B because of the following reasons:

\noindent $\bullet$ If the broad-H$\alpha$ originates from the central AGN, the X-ray luminosity of AGN can be evaluated by using the empirical correlations between the broad H$\alpha$ luminosity, continuum luminosity $L_{5100}$, and $L_{2-10{\rm{keV}}}$ \citep{Greene05,Jin12}, i.e., $\log L_{H\alpha}=-8.19+1.16\log L_{5100}$ and $\log L_{2-10{\rm{keV}}}=1.08\log L_{5100}-4.07$. 
The inferred $\log L_{2-10{\rm{keV}}}$ of AGN is $\sim\!42.2\pm 0.5$ (in units of $\log[{\rm erg/s}]$), much higher than the observation $\log L_{2-10{\rm{keV}}}<40.2$. Therefore, RGG\,B may not contain an AGN or the broad-H$\alpha$ component is not caused by the accretion disk of an AGN.

\begin{figure}
\centering
\includegraphics[width=\columnwidth]{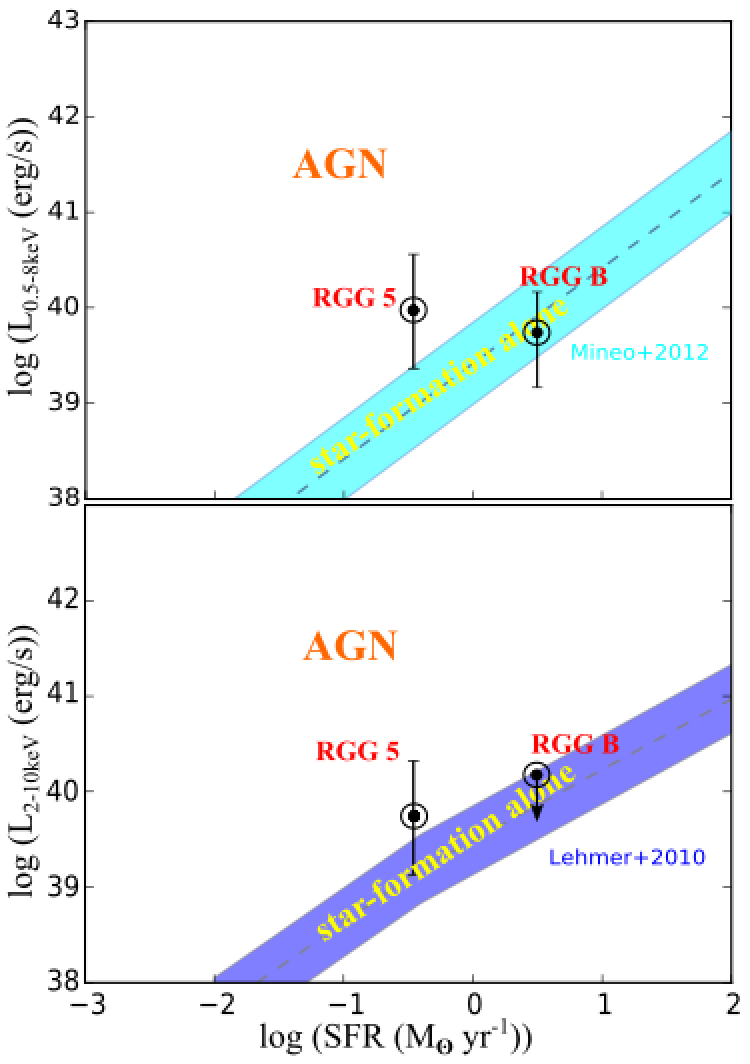}
\caption{The X-ray luminosity versus SFR for RGG\,B and RGG\,5. The X-ray luminosities of RGG~B are from Chandra observation Baldassare et al.~(2017); the arrow indicates the upper luminosity limit. The X-ray luminosities of RGG~5 are estimated from the NIR luminosities and typical AGN SED by assuming or the NIR light of the blueberry core is contributed by the AGN; the error bars of RGG\,5 are obtained by assuming the AGN SED in the study of Richards et al.~(2006) has a scatter of 0.5~dex.~Note that the predicated X-ray luminosities of RGG\,5 are the upper limits.~SFRs are determined based on the luminosity of H$\alpha$.~The upper panel uses 0.5-8~keV luminosities and the relation defined in Mineo et al.~(2012); the lower panel uses 2-10~keV luminosities and the relation defined by Lehmer et al. (2010).~The cyan and blue shaded regions show the relations for star formation alone.}
\label{SFR_Lx}
\end{figure}

\noindent $\bullet$ Its X-ray luminosity and SFR obtained from the narrow-H$\alpha$ luminosity \citep{Kennicutt98} follow the correlation between X-ray luminosities and SFRs expected for star formation alone \citep{Lehmer10,Mineo12,Baldassare17}, as shown in Fig.~\ref{SFR_Lx}, suggesting that the X-ray radiation of RGG\,B is more likely to be contributed by star formation.

\noindent $\bullet$ The ratio $\alpha_{\rm OX}\!=\!-0.383\log(l_{2500}/l_{2{\rm keV}})\!\simeq\!-2.13$ of RGG\,B, where $l_{2500}$ and $l_{2{\rm keV}}$ are the luminosity densities at $2500$\,\AA\ and 2\,keV, respectively,
is far off from the expected value from a quasar \citep{Just07,Baldassare17}.


\noindent $\bullet$ In general, the low-mass AGN in dwarf galaxies show high-metallicities similar to those of the typical high-mass AGN \citep{Greene07}, inconsistent with the low-metallicity of RGG\,B.


RGG\,5 shows no broad-line emission, or the S/N of the broad component is so low that we cannot extract such a weak component.~In the BPT diagram, RGG\,5 is located slightly beyond (excess of $\sim\!0.01$~dex) the theoretical maximum-starburst-line.~We also study its locations in the diagrams of [O\,{\sc iii}]/H$\beta$ versus [S\,{\sc ii}]/H$\alpha$ and [O\,{\sc iii}]/H$\beta$ versus [O\,{\sc i}]/H$\alpha$, and find that RGG\,5 always resides in the BPT AGN region and is slightly beyond (excess of $\lesssim 0.07$~dex) the theoretical maximum-starburst-lines.~Considering that there are about 0.1~dex errors of the theoretical maximum-starburst-lines \citep{Kewley01}, it definitely cannot exclude the possibility that RGG\,5 is a star-forming galaxy rather than AGN. Further, in the [O\,{\sc iii}]$_{\lambda5007}$/[O\,{\sc ii}]$_{\lambda3727}$ versus [O\,{\sc i}]$_{\lambda6300}$/H$\alpha$ diagram (panel~C of Fig.~\ref{property}), RGG\,5 is located at the H\,{\sc ii} rather than Seyfert region, suggesting that RGG\,5 is likely not hosting a significant AGN component.

Another valid method of judging whether RGG\,5 contains an AGN is to estimate its X-ray luminosity from another-passband (e.g., NIR) luminosity by using a typical AGN SED. Indeed, this method that derives one-passband luminosity from another-passband luminosity has been widely used and proved to be succesful in some other dwarf galaxies in the previous literatures \citep[e.g.,][]{Baldassare17a,Marleau17,Stern07}, although it introduces a relatively large uncertainty of $\sim0.5$~dex. If we assume that the F110W luminosity of the blueberry core (in Fig.~\ref{HST}) of RGG\,5 is dominated by AGN radiation, then using the AGN SED from \cite{Richards06}, we can infer the expected AGN X-ray luminosity from the observed F110W luminosity.~The SEDs of the optically-blue and IR-luminous AGN in the study of \cite{Richards06} are used, respectively, to estimate the X-ray luminosity. We obtain $L_{0.5-8{\rm{keV}}}\lesssim 1.15\times 10^{40}$~erg/s and $L_{2-10{\rm{keV}}}\lesssim 6.67\times 10^{39}$~erg/s for an optically blue AGN, and $L_{0.5-8{\rm{keV}}}\lesssim 7.38\times 10^{39}$~erg/s and $L_{2-10{\rm{keV}}}\lesssim 4.18\times 10^{39}$~erg/s for an IR-luminous AGN. Note, the SEDs of AGN in the study of \cite{Richards06} have a large scatter of $\sim 0.5$~dex. As explored in Fig.~\ref{SFR_Lx}, for RGG\,5, the predicted $L_{2-10{\rm{keV}}}$ versus SFR, and $L_{0.5-8{\rm{keV}}}$ versus SFR relations, are only slightly beyond the correlations between the X-ray luminosities and SFRs expected for star formation alone \citep{Lehmer10,Mineo12}; the error bars of RGG\,5 correspond to the $\sim\!0.5$~dex scatter of the AGN SEDs. However, we stress that the predicted X-ray luminosities are the upper limits of RGG\,5, as the F110W NIR luminosities of RGG\,5 should contain contributions from star formation. Therefore, the positions of RGG\,5 should be much closer to the region of star formation alone, thus RGG\,5 is not likely to harbor an AGN or only reveals a very weak and negligible AGN activity.

Here we conclude that the blueberry galaxies may not harbor AGNs. There are some possible reasons for non-existence of AGNs in the blueberries, e.g., the blueberries are in the early stages of merging where potential AGN activities have not been triggered \citep{Guo11,Rong17b}, or these blueberry galaxies do not contain massive black holes at all. Again, we note that the uncertainties of the predicted X-ray luminosities from the NIR luminosities and typical AGN SEDs are relatively large, we appeal to obtain further X-ray observations for more blueberry galaxies to test our conclusions.

\subsection{The case of borad-line emission}
As shown in Fig.~\ref{spectra}, RGG\,B presents broad-line emission (e.g., broad-H$\alpha$) corresponding to a large velocity of gas $\sim\!1000\!-\!2000$~km/s, which was believed to be caused by the supernovae (SNe) shocks or stellar winds from early-type stars in the previous studies \citep[e.g.,][]{Izotov07,Izotov08,Pustilnik04}. Generally, the broad emission in the low-metallicity BCDs with broad-H$\alpha$ luminosities of $10^{36}\--10^{40}\ \rm{ergs\ s^{-1}}$ is likely to arise from circumstellar envelopes around hot Ofp/WN9 and/or LBV stars, or from single or multiple SN remnants propagating in the interstellar medium, or from SN bubbles; whereas the broad emission with higher broad-H$\alpha$ luminosities, such as RGG\,B ($\sim\!5.4\!\times\!10^{40}\ \rm{ergs\ s^{-1}}$), is probably produced by the shocks propagating in the circumstellar envelopes of type\,{\sc II}n SNe or an AGN \citep{Izotov07,Izotov08,Guseva00,Smith11}.

For RGG\,B, its broad-line emission is very likely attributed to the shocks.~Indeed, \cite{OHalloran08} found that the SNe shocks, which are traced by the ratios of the 26~$\mu$m [Fe\,{\sc ii}] line and 12.8~$\mu$m [Ne\,{\sc ii}] line, play an important role in the deficit of polycyclic aromatic hydrocarbons in RGG\,B. RGG\,5 presents no broad-line emission; yet we cannot exclude the possibility that it is due to the low S/N ratio or that the SDSS fiber only covered the nucleus region while the broad-line emission is produced in the outskirts.~Particularly, the shell-like structures in RGG\,5 is proposed to be formed in the recent star formation regions in a shocked galactic wind \citep{Fabian80}. We suspect that the shocks may broadly exist in the blueberries. Note, the high [O\,{\sc iii}]/H$\beta$ of the blueberry galaxies may also be partly attributed to the boosting of shocks \citep[e.g.,][]{Guerrero08,Guerrero13}.

Compared with the typical BCDs, e.g.~Mrk~930, which presents the WR spectral features \citep{Adamo11b}, the blueberries are more ionized, indicating that their stellar populations are likely younger, and thus the WR stellar populations are expected to be found in blueberries. Additionally, the N overabundances of the blueberry galaxies may also suggest the non-negligible effect of the WR stars. Indeed, the WR blue bump and significant He\,{\sc ii}$_{\lambda4686}$ emission line have been verified in RGG\,B \citep{Pustilnik04}. However, in contrast to RGG\,B, RGG\,5 reveals no He\,{\sc ii}$_{\lambda4686}$ emission or WR blue bump (see Fig.~\ref{spectra}), which may be because of the limited S/N ratio of the spectrum, or the WR stars are primarily distributed in the outskirt of RGG\,5 while the nucleus is dominated by the red supergiants.~In this sense, high-S/N IFU spectroscopic observation with, e.g.~the Multi Unit Spectroscopic Explorer (MUSE) on ESO/VLT, would be ideal to characterize the properties of the WR stars in RGG\,5.

\section{Summary}\label{sec:4}

This is a pilot study of blueberry galaxies in the nearby Universe.~We have studied the spectroscopic and photometric properties of two blueberry candidate galaxies, RGG\,B and RGG\,5, with high-resolution HST images and SDSS spectroscopy.~We found that their properties, e.g., BPT locations, high-sSFRs, low-metallicities, high-ionization ratios, nitrogen-overabundances, and mid-IR colors, are similar to those of the Y17 blueberries, and thus can be typical examples to understand the characteristics of blueberries.~In the following, we summarize some key findings.

1) The blueberry galaxies perhaps are merging dwarf galaxies, and contain luminous H\,{\sc ii} regions (possibly dominated by red supergiants) and some star-forming hotspots harboring many young star clusters.~Yet we cannot exclude the possibility that some blueberries may not undergo a merging event but simply exhibit one/several star-forming regions with outflows.~Future IFU observations will help resolving this issue.

2) All of the blueberries are approximately located at the upper-left star-forming region in the BPT diagram, and very close to the theoretical maximum-starburst-line. All of the blueberries have very high ionization parameters and relatively low hardness ionizing radiation field, show N overabundances and extremely red mid-IR colors and reside in the so-called ``ULIRGs/LINERs/Obscured AGN'' region. The red mid-IR colors of the blueberries are probably caused by the UV photons from the young massive star clusters in the galaxies heating the surrounded dust.

3) According to the locations in the BPT and SFR versus X-ray luminosity diagrams, the blueberry galaxies may not harbor AGNs. We appeal to collect deeper X-ray observations for more blueberries to test this possible conclusion.

\section*{Acknowledgments}

YR thanks Simon Pustilnik, Chen Hu for their helpful discussions, and acknowledges supports from CAS-CONICYT postdoctoral fellowship No.\,16004 and NSFC grant No.\,11703037. This work is accomplished with the support from the Chinese Academy of Sciences (CAS) through a CAS-CONICYT Postdoctoral Fellowship administered by the CAS South America Center for Astronomy
(CASSACA) in Santiago, Chile. THP acknowledges support in form of the FONDECYT Regular Project No.\,1161817 and by the BASAL Center for Astrophysics and Associated Technologies (PFB-06).

\bibliographystyle{mn2e}


\end{document}